\documentclass[11pt]{article}
\usepackage{geometry}
\usepackage{amsmath, amssymb}
\usepackage[retainorgcmds]{IEEEtrantools}
\usepackage[all,knot,poly]{xy}
\usepackage{rotating}

\usepackage{graphicx}	
\usepackage{float}
\usepackage[utf8]{inputenc}
\usepackage{multirow}
\usepackage{array}
\usepackage{mathtools}
\usepackage{makecell}
\usepackage{diagbox}
\usepackage{soul}
\usepackage{color}
\numberwithin{equation}{section}
\usepackage{multicol,caption}
\usepackage{empheq}
\usepackage{enumitem}
\usepackage{verbatim}
\usepackage{microtype}
\usepackage[super,sort&compress]{natbib} 

\title{The Concept of Electric Charge and\\ the Hypothesis of Magnetic Poles}
\author{Robert J. Finkelstein}
\date{\emph{Physics \& Astronomy, University of California, Los Angeles, \\
475 Portola Plaza, Los Angeles, California 90095, USA \\
finkel@physics.ucla.edu}}

\begin{document}

\setlength{\baselineskip}{1.6\baselineskip}

\maketitle

\begin{abstract}
\setlength{\baselineskip}{1.6\baselineskip}

We examine a generic field theory in which the field particle has two couplings. It is of particular interest when these are the electroweak, e, and the hypothetical magnetoweak, g. The new field operators are obtained by replacing the field operators $\Psi (x)$ of the standard model or of similar models by $\tilde{\Psi} (x) D^j_q (m,m')$ where $ D^j_q (m,m')$ is an element of the $2j+1$ dimensional representation of the SLq(2) algebra, which is also the knot algebra. The new field is assumed to exist in two phases distinguished by two values of $q$: $q_e = \frac{e}{g}$ and $q_g = \frac{g}{e}$ which label the electroweak and magnetoweak phases respectively. 

We assume that the observed leptons and quarks are mainly composed of e-preons and are in agreement with the observed charge spectrum of leptons and quarks. It is now proposed that there is also a g-phase where g-leptons and g-quarks are composed of mainly g-preons. It is assumed that the g-charge is very large compared to the e-charge and the mass of the g charged particle is even larger since the mass of all of these particles is partially determined by the eigenvalues of $\bar{D}^j_q (m,m') D_q^j (m,m')$, a polynomial in $q$, that multiplies the Higgs mass term and where
\begin{equation*}
\frac{q_g}{q_e} =  \left ( \frac{\hbar c}{e^2} \right)^2 \approx (137)^2.
\end{equation*}

Since these values of $q$ indicate that particles in the g-phase are much more massive, they should be harder to produce or to observe. Since the remote parts of the universe are at increasingly higher temperatures, magnetic poles are perhaps most likely to be found in deep probes of space as well as in high energy accelerators.

The section entitled "Introduction" was added only after it was generally realized that the birth of the present universe was probably a nuclear explosion.

\end{abstract}

\section{Introduction}

When Maxwell in 1810 combined the experimental discoveries of Faraday on the separate electric and magnetic fields into a unified presentation of the electromagnetic field, he chose to include electric charges but not magnetic monopoles, since at the time there was no evidence of the latter. This omission left a theoretical question which Dirac addressed in 1911, and again in 1948, and was further discussed by Schwinger in 1969 who hypothesized dyons that carried both electric and magnetic charge.
\paragraph{} Although the Quantized Maxwell theory is a highly successful theory, its unquantized version also permitted an interpretation as a theory of Light (with the correct and already known velocity of Light) but with respect to an unknown reference frame, which came to be called the "Universal Ether" and was supposed to fill all of space. Next came the null result of the Michelson-Morley experiment followed by Einstein's proposed (1905) special theory of relativity which dispensed with the "ether". Then in 1910 came General Relativity which provided a way to compare with the fairly accurate Newtonian equations for planetary orbits. More importantly, General Relativity established space-time as the "unknown reference frame" implied by Maxwell's equations.
\paragraph{} As generally believed at that time, and as also assumed by Einstein in applying his field equations with Schwartzchild boundary conditions, the universe was assumed to be static. By 1929, however, Hubble had shown, by careful examination of the visible red shifts of the visible galaxies, that the universe was expanding. Only one theorist, Alexander Friedmann, a Russian mathematical physicist based in Moscow, had examined the Einstein field equations under the following different boundary conditions\cite{hawking}:
\begin{enumerate}
    \item The universe looks identical in whatever direction one looks
    \item The same would be true if one were observing from anywhere else
\end{enumerate}
It then turned out that Friedmann had predicted exactly what Hubble had found. Friedmann's work was rediscovered and confirmed in 1935 by the American and British physicists, Howard Robertson and Arthur Walker and is also known today as the Robertson-Walker form of the Einstein field equations.
\paragraph{} Much later, in 1965, Arno Penzias and Robert Wilcox at Bell Telephone Laboratories in New Jersey, detected mysterious microwave radiation, which later turned out to be a remarkably accurate confirmation of Friedmann's assumptions and calculations. 
\paragraph{} At nearly the same time as Penzias and Wilson, two members of the Princeton Faculty, Robert Dicke and James Peebles, were exploring a suggestion of George Gamow (who had been a student of Friedmann), that the Friedmann Universe was the result of a nuclear explosion and therefore very hot and dense, and that they might be able to see its glow because light from it would still be reaching the earth in 1964. They were preparing to look for a greatly red-shifted microwave radiation produced at the birth of the expanding Hubble-Friedmann universe when they realized that Penzias and Wilson had probably already identified the mysterious radiation that had puzzled them.
\paragraph{} The relatively recent Gamow scenario relating the Friedmann solution of Einstein field equations under appropriate boundary conditions to the state of the universe in 1964 and subsequent years was confirmed by the observation of the red-shifted microwave data with wavelength corresponding to about 2.7 Kelvin.
\paragraph{} We are mainly interested here in whether the evidence for magnetic poles is thereby strengthened, since magnetic poles would be present in the early universe according to the scenario presented in this paper.
\paragraph{} In recent papers we have proposed a topological description of electroweak charge.\cite{finkelstein14a}$^,$\cite{finkelstein14b}$^,$\cite{finkelstein15} By an extension of the same considerations one is led to consider magnetic charge and the existence of magnetic poles. We begin with the remark that the standard model of the elementary particles depends on two couplings, $e$ and $g$, and in addition incorporates the electroweak theory that also depends on two couplings, there related by the Weinberg angle. In a third theoretical model, suggested by Schwinger, the elementary particles are called dyons and are sources of both electric and magnetic charge.\cite{schwinger} Here we shall describe some aspects of a generic field theory where the field quanta have two couplings that may be expressed in the coupling matrix
\begin{equation}
\boxed{
\varepsilon_q = \begin{pmatrix}
0 & \alpha_2 \\
-\alpha_1 & 0
\end{pmatrix}.} \label{couplingmatrix}
\end{equation}
The couplings $\alpha_1$ and $\alpha_2$ are assumed to be dimensionless and real and may be written as
\begin{equation}
(\alpha_1, \alpha_2) \text{ or } (\alpha_2, \alpha_1) = \left( \frac{e}{\sqrt{\hbar c}}, \frac{g}{\sqrt{\hbar c}} \right) \label{couplingconstants}
\end{equation}
where $e$ and $g$ refer to a specific two charge model and have dimensions of an electric charge. We assume that e and g may be energy dependent and normalized at relevant energies. The reference charge is the universal constant $\sqrt{\hbar c}$. 

The fundamental assumption that we make on this coupling matrix is that it is invariant under SLq(2) as follows
\begin{equation}
T \varepsilon_q T^t = T^t \varepsilon_q T = \varepsilon_q \label{slq2invariant}
\end{equation}
where $t$ means transpose and $T$ is a two dimensional representation of SLq(2):
\begin{equation}
T = \begin{pmatrix}
a & b \\
c & d
\end{pmatrix} . \label{2drepslq2}
\end{equation}
By \eqref{couplingmatrix}, \eqref{slq2invariant} and \eqref{2drepslq2} the elements of $T$ obey the knot algebra:
\begin{equation}
\begin{split}
ab = qba \qquad bd = qdb \qquad ad-qbc = 1 \qquad bc &= cb \\
ac = qca \qquad cd = qdc \qquad da -q_1cb = 1 \qquad q_1 &\equiv q^{-1}
\end{split}
\tag{A}
\end{equation}
where
\begin{equation}
q = \frac{\alpha_1}{\alpha_2}
\end{equation}
so that the two physical couplings fix the algebra through their ratio.

If also
\begin{equation}
\text{det} \hspace{2pt} \varepsilon_q = 1 \label{couplingdet}
\end{equation}
one has
\begin{equation}
\alpha_1 \alpha_2 = 1
\end{equation}
If the two couplings $(\alpha_1, \alpha_2)$ are given by \eqref{couplingconstants}, where $e$ and $g$ are the electroweak and ``gluon''-like couplings, or electric and magnetic couplings, then
\begin{equation}
eg = \hbar c \label{quantize}
\end{equation}
Then \eqref{couplingconstants} and the normalizing condition \eqref{couplingdet} imply that $q_g$ is the reciprocal of the fine structure constant:
\begin{IEEEeqnarray}{*x+rCl+}
 &q_g & =& \frac{g}{e} \\
\text{then}& q_g&  =& \frac{\hbar c}{e^2} \sim 137
\end{IEEEeqnarray}
If $g$ represents magnetic charge, then \eqref{quantize} resembles the Dirac requirement according to which the magnetic charge is very much stronger than the electric charge. If the Dirac magnetic pole is very much heavier as well, it may be observable only at early and not at current cosmological temperatures or at currently achievable accelerator energies.

We shall assume that magnetic poles do exist and shall study the possible extension of knot symmetry to magnetic charges.

\section{Topological Constraints\cite{finkelstein14a}$^,$\cite{finkelstein14b}$^,$\cite{finkelstein15}}

In the models considered here we replace the quantum field operators $\Psi (x)$ of the standard model, or of similar models, by $\tilde{\Psi} ^j_q (m, m') D^j_q (m, m')$ where $\tilde{\Psi} ^j_q (m, m')$ satisfies the standard Langrangian or a similar Langrangian after modification by the form factors generated by the $D^j_q (m,m')$. The $D^j_q (m,m')$ partially label the quantum states of the field particles. We postulate that these states are restricted by the topological spectrum of a corresponding classical knot as follows:
\begin{equation}
\boxed{
(j,m,m')_q = \frac{1}{2}(N, w, r+o) \label{knotrestriction}}
\end{equation}
where $N$, $w$, and $r$ are respectively the number of crossings, the writhe, and the topological rotation of the 2d projection of a corresponding classical knot. Here $o$ is an odd number required by the otherwise forbidden difference in parity between the two sides of \eqref{knotrestriction}. Eqn.~\eqref{knotrestriction} then describes a correspondence between quantum knots $(j,m,m')_q$ and the 2d-projected classical knots $(N, w, r)$. The dynamical evolution of the field is described by the quantum field theory but this evolution is kinematically constrained by the classical knot topology as expressed in \eqref{knotrestriction}. The odd number o will be set at unity for the simplest knot, the trefoil.

\section{Knot Polynomials\cite{finkelstein14b}}
As given by \eqref{couplingmatrix} and \eqref{couplingdet}, $\varepsilon_q$ reduces to
\begin{equation}
\varepsilon_q = \begin{pmatrix}
0 & q^{-\frac{1}{2}} \\
-q^{\frac{1}{2}} & 0
\end{pmatrix}
\end{equation}
This matrix, together with the Pauli matrices, underlies the structure of the Kauffmann knot polynomial,\cite{finkelstein15} which is a Laurent polynomial in $q$. Here we do not in general assume \eqref{couplingdet} and are interested in different knot polynomials, namely the $2j+1$ dimensional representations of SLq(2) acting on the following Weyl monomial basis.\cite{finkelstein15}
\begin{equation}
\Psi_m^j = N_m^j x^{n_+ (m)}_1 x_2^{n_- (m)} \qquad \qquad -j \leq m \leq j
\end{equation}
where $x_1$ and $x_2$ are coordinates on a privileged plane that may also be used to define parity. Here
\begin{equation}
[x_1, x_2] = 0
\end{equation}
and
\begin{equation}
n_{\pm} (m) = j \pm m
\end{equation}

The normalizing factor is
\begin{equation}
N_m^j = [\langle n_+ (m) \rangle_{q_1}! \langle n_- (m) \rangle_{q_1} ! ]^{-\frac{1}{2}}
\end{equation}
where
\begin{equation}
q_1 = q^{-1}
\end{equation}
and
\begin{equation}
\boxed{\langle n \rangle_q = \frac{q^n -1}{q-1} \label{nq}}
\end{equation}
\emph{Then if T is the two-dimensional representation of SLq(2) as expressed in {(1.4)} and,}
\begin{equation}
\begin{pmatrix}
x_1 \\
x_2
\end{pmatrix}'
= T \begin{pmatrix}
x_1 \\ x_2
\end{pmatrix}
\end{equation}
\emph{Eqn. (3.8) induces}
\begin{equation}
\Psi^{j'}_m = D^j_{mm'} \Psi^j_{m'}
\end{equation}
where
\begin{equation}
\boxed{ {D^j_{mm'} = \sum_{n_a, n_b, n_c, n_d} A(q | n_a, n_b, n_c, n_d) a^{n_a} b^{n_b} c^{n_c} d^{n_d} \label{long}}}
\end{equation}

Here $a, b, c, d$ satisfy the algebra (A) and the sum  on $n_a, n_b, n_c, n_d$ is over all positive integers and zero that satisfy the following equations: \cite{finkelstein14a}
\begin{empheq}[box=\fbox]{align}
    n_a + n_b + n_c + n_d &= 2j \label{n2j}\\
    n_a + n_b - n_c - n_d &= 2m \label{n2m}\\
    n_a - n_b + n_c - n_d &= 2m' \label{n2m'}
\end{empheq}

and
\begin{equation}
\boxed{A (q \vert n_a n_b n_c n_d) = \left [ \frac{ \langle n_+ ' \rangle_1 ! \langle n_- ' \rangle_1 ! }{\langle n_+ \rangle_1 ! \langle n_- \rangle_1 !} \right ]^{\frac{1}{2}} \frac{ \langle n_+ \rangle_1 !}{\langle n_a \rangle_1 ! \langle n_b \rangle_1 !} \frac{\langle n_- \rangle_1 !}{\langle n_c \rangle_1 ! \langle n_d \rangle_1 !}}
\end{equation}

The two dimensional representation, $T$, introduced by \eqref{2drepslq2} now reappears as the $j = \frac{1}{2}$ fundamental representation
\begin{equation}
\boxed{D^{\frac{1}{2}}_{mm'} = \begin{pmatrix} a & b \\ c & d \end{pmatrix} \label{jhalfrep}}
\end{equation}
In the physical model with the \eqref{couplingmatrix} coupling we interpret $(a,b,c,d)$ in \eqref{long} as creation operators for $(a,b,c,d)$ particles, which we term preons. Then $D^j_{mm'}(a,b,c,d)$ is the creation operator for the state representing the superposition of $(n_a, n_b, n_c, n_d)$ preons. Since we shall regard the preons as fermions, they will also carry an anti-symmetrizing index to satisfy the Pauli principle.

\section{Noether Charges carried by $D^j_{mm'}$ knots\cite{finkelstein14b}}
The knot algebra (A) is invariant under
\begin{eqnarray}
a' = e^{i \varphi_a} a & \quad & b' = e^{i \varphi_b} b \\
d' = e^{-i \varphi_a} d & \quad & c' = e^{-i \varphi_b} c \nonumber
\end{eqnarray}
We shall refer to the transformation described by (4.1) as $U_a (1) \times U_b (1)$.

The transformation, $U_a (1) \times U_b (1)$, on the $(a,b,c,d)$ of SLq(2) induces on the $D^j_{mm'}$ of SLq(2) the corresponding transformation\cite{finkelstein15}
\begin{IEEEeqnarray}{rCl}
D^j_{mm'} (a,b,c,d) & \rightarrow & D^j_{mm'} (a',b',c',d') \\
& = & e^{i(\varphi_a + \varphi_b)m}e^{i(\varphi_a - \varphi_b)m'}D^j_{mm'}(a,b,c,d) \\
&= & U_m(1) \times U_{m'}(1)D^j_{mm'}(a,b,c,d)
\end{IEEEeqnarray} 
and on the field operators as modified by the $D^j_{mm'}$
\begin{equation}
\Psi^j_{mm'} \rightarrow U_m(1) \times U_{m'}(1) \Psi^j_{mm'} \label{fieldeigen}
\end{equation}

\emph{For physical consistency any knotted field action must be invariant under \eqref{fieldeigen} since \eqref{fieldeigen} is induced by $U_a \times U_b$ transformations that leave the defining algebra (A) unchanged.}
\emph{There are then Noether charges associated with $U_m$ and $U_{m'}$ that may be described as writhe and rotation charges, $Q_w$ and $Q_r$, since $m=\frac{w}{2}$ and $m'=\frac{1}{2}(r+o)$ for quantum knots.}

For quantum trefoils we have set $o=1$, and we now define their Noether charges:
\begin{empheq}[box=\fbox]{align}
    Q_w \equiv -k_wm \equiv -k_w\frac{w}{2} \label{qwdef} \\
    Q_r \equiv -k_rm' \equiv -k_r\frac{1}{2}(r+1)  \label{qrdef}
\end{empheq}

where $k_w$ and $k_r$ are undetermined charges.

The generic model based on $D^{N/2}_{\frac{w}{2} \frac{r+1}{2}}$ has been worked out in some detail as a SLq(2) extension of the standard lepton-quark model at the electroweak level.\cite{finkelstein14a}$^,$\cite{finkelstein14b}$^,$\cite{finkelstein15} It is surprisingly successful when formulated as a preon theory. Being a new model, however, it presents some unanswered questions and in particular it does not predict whether the preons are bound or are in fact observable. Since the hypothetical preons are presumably much smaller and heavier than the leptons and quarks, a very strong binding force is required to permit one to regard the leptons and quarks as composed of three observable preons. The binding force could be gravitational and it could also be dyonic as suggested by Schwinger. To study the gravitational model it might be necessary to examine the possibility that space-time at earliest times, or highest energies, becomes 2+1 dimensional.\cite{carlip} To study the dyonic model one could assume that the preons are dyons. In both the gravitational and dyonic models one is exploring very high energies or very early cosmological times and in both cases extremely microscopic structure, while the standard model has been calibrated and tested only at the electroweak level and at accessible accelerator energies. 

The rest of this section and the following sections (5--13) repeat earlier work that refers explicitly to the e-phase. It is repeated here as compatible with the invariance of the coupling matrix \eqref{couplingmatrix} and it therefore also provides a possible description of the g-phase.

The question that we wish to examine here is whether there is a formulation of the SLq(2) algebra such that the preon formulation of the electroweak model can be reinterpreted and reparameterized at higher energies to realistically also describe a dyonic Lagrangian of observable dyons.

To approach this question we summarize the SLq(2) extension of the standard model by first restricting the states described by the field operators, $\tilde{\Psi}^j_q (m, m') D^j_q (m, m')$, to states obeying the postulated relations \eqref{knotrestriction}

\begin{equation}
\centering
\boxed{
(j,m,m')_q = \frac{1}{2}(N, w, r+o)} \label{jN}
\end{equation}

and also the \textit{empirically based relations} that appear in Table 1\cite{finkelstein14a}$^,$\cite{finkelstein14b}$^,$\cite{finkelstein15}

\begin{IEEEeqnarray}{*x+rCl+}
\centering
\text{namely,}
\Aboxed{& (N,w,r+1)  =  6 (t, -t_3, -t_0)} \\ [8pt]
\text{or by (4.8) and (4.9)} & \Aboxed{(j, m, m')_q  = 3(t,-t_3,-t_0)} \label{empirical}
\end{IEEEeqnarray}

Eqn.~(4.9) holds for $j = \frac{3}{2}$ and $t = \frac{1}{2}$ as shown in Table 1. Since $t$ and $t_3$ refer to isotopic spin and $t_0$ refers to hypercharge, Table 1 describes a \textit{correspondence between the simplest fermions and the simplest knots.}

\renewcommand{\arraystretch}{2}

\begin{figure}[H]
\begin{tabular}{r c r r r | p{14mm} c r r c | r}
\multicolumn{10}{c}{\textbf{Table 1:} Empirical Support for $6(t,-t_3,-t_0) = (N,w,r+1)$} \\ 
\hline \hline
 & Elementary Fermions & $t$ & $t_3$ & $t_0$ & Classical Trefoil & $N$ & $w$ & $r$ & $r+1$ & $D^{N/2}_{\frac{w}{2}\frac{r+1}{2}}$ \\[0.2cm]
\hline
\multirow{2}{*}{\hspace{-4pt}leptons $\Bigg \{$} & $(e, \mu, \tau)_L$ & $\frac{1}{2}$ & $-\frac{1}{2}$ & $-\frac{1}{2}$ & & 3 & 3 & 2 & 3  & $D^{3/2}_{\frac{3}{2}\frac{3}{2}}$\\
& $(\nu_e, \nu_{\mu}, \nu_{\tau})_L$ & $\frac{1}{2}$ & $\frac{1}{2}$ & $-\frac{1}{2}$ & & 3 & $-3$ & 2 & 3 & $D^{3/2}_{-\frac{3}{2}\frac{3}{2}}$\\
\multirow{2}{*}{quarks $\Bigg \{$} & $(d, s, b)_L$ & $\frac{1}{2}$ & $-\frac{1}{2}$ & $\frac{1}{6}$ & & 3 & 3 & $-2$ & $-1$ & $D^{3/2}_{\frac{3}{2}-\frac{1}{2}}$\\
& $(u, c, t)_L$ & $\frac{1}{2}$ & $\frac{1}{2}$ & $\frac{1}{6}$ & & 3 & $-3$ & $-2$ & $-1$ & $D^{3/2}_{-\frac{3}{2}-\frac{1}{2}}$\\ [0.2cm]
\hline
\end{tabular}

\caption*{\textit{The symbols $(\quad)_L$ designate the left chiral states in the usual notation. The topological labels $(N,w,r)$ on the right side provide a natural way to abstractly label the same chiral states.}}
\end{figure}

Then the $D^{N/2}_{\frac{w}{2}\frac{r+1}{2}}$ describe the allowed states of the quantum trefoil.

Only for the particular row-to-row correspondences shown in Table 1 does \eqref{empirical} hold, i.e., \emph{each of the four families of fermions labelled by $(t_3, t_0)$ is uniquely correlated with a specific $(w,r)$ classical trefoil, and therefore with a specific state $D^{N/2}_{\frac{w}{2}\frac{r+1}{2}}$ of the quantum trefoil.}

Note that the $t_3$ doublets of the standard model now become the writhe doublets ($w = \pm 3$). Note also that with this same correspondence the leptons and quarks form a knot rotation doublet ($r = \pm 2$).

Retaining the row to row correspondence described in Table 1, it is then possible to compare in Table 2 the electroweak charges, $Q_e$, of the most elementary fermions with the total Noether charges, $Q_w + Q_r$, of the simplest quantum knots, which are the quantum trefoils.

\begin{center}
\begin{tabular} {c r r r c | c l c c c}
\multicolumn{10}{c}{{\textbf{Table 2:} Electric Charges of Leptons, Quarks, and Quantum Trefoils}} \\
\hline \hline
\multicolumn{5}{c |}{{Standard Model}} & \multicolumn{5}{c}{{Quantum Trefoil Model}} \\
\hline
{$(f_1, f_2, f_3)$} & {$t$} & {$t_3$} &{$t_0$} & {$Q_e$} & {$(N,w,r)$} & {$D^{N/2}_{\frac{w}{2}\frac{r+1}{2}}$} & {$Q_w$} & {$Q_r$} & {$Q_w + Q_r$} \\ [0.2cm]
\hline
$(e, \mu, \tau)_L$ & $\frac{1}{2}$ & $-\frac{1}{2}$ & $-\frac{1}{2}$ & $-e$ & $(3,3,2)$ & $D^{3/2}_{\frac{3}{2} \frac{3}{2}}$ & $-k_w \left( \frac{3}{2} \right)$ & $-k_r \left( \frac{3}{2} \right)$ & $-\frac{3}{2}(k_r + k_w)$ \\
$(\nu_e, \nu_{\mu}, \nu_{\tau})_L \hspace{-10pt}$ & $\frac{1}{2}$ & $\frac{1}{2}$ & $-\frac{1}{2}$ & $0$ & $(3,-3,2)$ & $D^{3/2}_{-\frac{3}{2} \frac{3}{2}}$ & $-k_w \left( -\frac{3}{2} \right)$ & $-k_r \left( \frac{3}{2} \right)$ & $\frac{3}{2}(k_w - k_r) $ \\
$(d,s,b)_L$ & $\frac{1}{2}$ & $-\frac{1}{2}$ & $\frac{1}{6}$ & $-\frac{1}{3}e$ & $(3,3,-2)$ & $D^{3/2}_{\frac{3}{2} -\frac{1}{2}}$ & $-k_w \left( \frac{3}{2} \right)$ & $-k_r \left( -\frac{1}{2} \right)$ & $\frac{1}{2}(k_r - 3k_w)$ \\
$(u,c,t)_L$ & $\frac{1}{2}$ & $\frac{1}{2}$ & $\frac{1}{6}$ & $\frac{2}{3}e$ & $(3,-3, -2)$ & $D^{3/2}_{-\frac{3}{2} -\frac{1}{2}} \hspace{-5pt}$ & $-k_w \left( -\frac{3}{2} \right)$ & $-k_r \left( -\frac{1}{2} \right)$ & $\frac{1}{2}(k_r + 3k_w)$ \\
& & \multicolumn{3}{c |}{$ \hspace{-8pt} Q_e = e(t_3+t_0) \hspace{-10pt}$} & \multicolumn{2}{c}{\normalsize $(j, m, m') = \frac{1}{2}(N, w, r + 1)$} & $Q_w = -k_w \frac{w}{2} \hspace{-5pt}$ & $Q_r = -k_r \frac{r+1}{2} \hspace{-15pt}$ & \\
\hline
\end{tabular}
\end{center}

 \emph{One sees that  \boxed{Q_w + Q_r = Q_e} is satisfied for charged leptons, neutrinos and for both up and down quarks with only a single value of $k$:}

\begin{equation}
k_r = k_w (= k)=\frac{e}{3} \label{3k}
\end{equation}
and also that $t_3$ isospin and $t_0$ hypercharge then measure the writhe and rotation charges respectively:
\begin{equation}
Q_w = et_3 \label{qwet3}
\end{equation}
\begin{equation}
Q_r = e t_0 \label{qret0}
\end{equation}
Then $Q_w + Q_r = Q_e$ becomes by \eqref{qwet3} and \eqref{qret0} an alternative statement of
\begin{equation}
Q_e = e(t_3 + t_0)
\end{equation}
of the standard model.

In SLq(2) measure $Q_e = Q_w + Q_r$ is by \eqref{qwdef} and \eqref{qrdef}:
\begin{equation}
\boxed{Q_e = -\frac{e}{3}(m+m'), \label{qeem}}
\end{equation}
or
\begin{equation}
Q_e = - \frac{e}{6}(w+r+1). \label{qeewr}
\end{equation}
for the quantum trefoils, that represent the elementary fermions.

Then the so defined electroweak charge is a measure of the writhe $+$ rotation of the trefoil. The total electroweak charge in this way resembles the total angular momentum as a sum of two parts where the knot rotation corresponds to the orbital angular momentum and where the localized contribution of the writhe to the charge corresponds to the localized contribution of the spin of the particle to the angular momentum, i.e. the writhe and topological rotation corresponds to ordinary spin and angular momentum respectively. In \eqref{qeewr} $o=1$ contributes a ``zero-point charge", similar to the energy of the lowest state of the harmonic oscillator. 

The total SLq(2) charge sums the signed clockwise and counterclockwise turns that any energy-momentum knotted current makes at the crossings and in one circuit of the 2d-projected knot. In this way, the ``handedness" or chirality of the knot determines its charge, so that knot chirality expresses electroweak charge as a geometrical concept similar to the way that curvature of space-time geometrizes mass and energy. This measure of charge, which is suggested by the charges of leptons and quarks, goes to a deeper level than the standard electroweak isotopic measure that was first suggested by the approximate equality of masses in the neutron-proton system. In the knot or topological definition of electroweak charge, chirality appears naturally with contributions from both the writhe, $t_3$, and the topological rotation, $t_0$. What had previously been ``hypercharge" is now simply topological rotation $(r)$ and what had previously been a new coupling is now a displaced chiral coupling. 

As here defined, quantum knots carry the charge expressed as both $t_3 + t_0$ and $m + m'$. The conventional \textbf{$\bf{(t_3, t_0)}$ measure of charge is based on $\bf{SU(2) \times U(1)}$ while the $\bf{(m,m')}$ measure of charge is based on SLq(2)}. These two different exact measures are related at the $j=\frac{3}{2}$ level by eqn.~(4.10): $(j,m,m')_q = 3(t,-t_3,-t_0)$.

We next extend this analysis beyond $j=\frac{3}{2}$, and in particular to the fundamental representation $j=\frac{1}{2}$. This extension to other states of $j$ is here intended to include as well a specialization of the generic model $(\alpha_1, \alpha_2)$ to $(e,g)/\sqrt{\hbar c}$ where $e$ is the electroweak coupling and $g$ is the ``magnetoweak'' coupling.
We shall repeat here a description of the electroweak phase. As far as we now know, a magnetoweak phase may be described along the same lines with $e$ replaced by $g$ in \eqref{3k}. Then the dyonic model would predict that the currently missing experimental support of the g-phase may have existed at earlier cosmological times. One may tentatively assume that there is a g-phase and that the $g$-particles have both $e$ and $g$ charges with $g >> e$, following Dirac.

We continue with the extension of \eqref{qeem} from $j = \frac{3}{2}$ to $j = \frac{1}{2}$ where the Noether charge is by (3.15) and \eqref{qeem}
\begin{equation}
Q_e = -\frac{e}{3}(m+m') \tag{4.15}
\end{equation}

\emph{By \eqref{jhalfrep}, (4.6), (4.7) and (4.15) there is one charged preon, a, with charge $-\frac{e}{3}$ and its antiparticle, d, and there is one neutral preon, b, with its antiparticle, c.}

If $j = \frac{1}{2}$, then $N=1$ by the postulate \eqref{knotrestriction} relating to the corresponding classical knot
\begin{equation}
    \boxed{(j,m,m') = \frac{1}{2}(N, w, r+o)} \tag{4.8}
\end{equation}
The corresponding $a,b,c,d$ classical pictures of the preons cannot therefore be described as knots since they have only a single crossing. They can, however, be described as \emph{2d-projections of twisted loops with $N=1$, $w= \pm 1$ and $r=0$}. 

Having tentatively interpreted the fundamental representation in terms of preons, we next consider the general representation.

\section*{Interpretation of all $D_{mm'}^j (q \vert a,b,c,d)$}
\emph {Every $D^j_{mm'}$, as given in \eqref{long}}

\emph{being a polynomial in $a,b,c,d$, can be interpreted as a creation operator for a superposition of states, each state with $n_a, n_b, n_c, n_d$ preons.} 

It then turns out that the creation operators \textbf{(} for the charged leptons, $D^{3/2}_{\frac{3}{2} \frac{3}{2}}$; neutrinos, $D^{3/2}_{-\frac{3}{2} \frac{3}{2}}$; down quarks, $D^{3/2}_{\frac{3}{2} -\frac{1}{2}}$; and up quarks, $D^{3/2}_{-\frac{3}{2} -\frac{1}{2}}$ \textbf{)}, \emph{as empirically required by Tables 1 and 2}, are represented by \eqref{long} as the following \textit{monomials, not polynomials}

\begin{center}
\begin{tabular}{c c c c c}
$D^{3/2}_{\frac{3}{2} \frac{3}{2}} \sim a^3$ & $D^{3/2}_{-\frac{3}{2} \frac{3}{2}} \sim c^3$ & $D^{3/2}_{\frac{3}{2} -\frac{1}{2}} \sim ab^2$ & $D^{3/2}_{-\frac{3}{2} -\frac{1}{2}} \sim cd^2$ & \hspace{22pt} (4.17) \\
\text{charged leptons} & \text{neutrinos} & \text{down quarks} & \text{up quarks}
\end{tabular}
\end{center}
implying that \emph{charged leptons and neutrinos are composed of three $a$-preons and three $c$-preons, respectively, while the down quarks are composed of one $a$- and two $b$-preons, and the up quarks are composed of one $c$- and two $d$-preons,  in agreement with the Harari,\cite{harari79} Shupe\cite{shupe79}and Raitio\cite{raitio} models, and with the experimental evidence on which their models are constructed.} Note that the number of preons equals the number of crossings ($(j = \frac{N}{2} = \frac{3}{2})$ in (4.17)).
\newline

Note also that there are only four ``elementary fermions" differing by the two possibilities for the writhe and the two possibilities for the rotation of the quantum trefoil. Each of the ``elementary fermions" has 3 states of excitation, determined by eigenstates of $\bar{D}^{3/2}_{mm'} D^{3/2}_{mm'}$\cite{finkelstein14a}, which are polynomials in $q$ that predict higher masses for the $q_g$ particles.

The discussion up to this point, and in the following part of this section as well, identifies the Noether charge with the electroweak charge in conformity with the empirical tables 1 and 2. Since the corresponding empirical support for a physical g sector has not yet been found, we are speculating that the g-phase has not yet been observed because it lies at a higher energy that is so far unobservable, and that the currently observable universe is an e-phase of the dyon field.

\section*{The SU(3) Couplings of the Standard Model}

The previous considerations are based on electroweak physics. To describe the strong interactions it is necessary according to the standard model to introduce SU(3). In the SLq(2) electroweak model, as here described, the need for the additional SU(3) symmetry is built into the knot model already at the level of the charged leptons and the neutrinos since they are presented as $a^3$ and $c^3$, respectively. \emph{Then the simple way to protect the Pauli principle is to replace} $(a,c)$ \emph{by} $(a_i, c_i)$ \emph{and antisymmetrize}

\begin{center}
\begin{tabular}{r l r}
\emph{the creation operators for charged leptons} & $a^3 \text{ by } \varepsilon^{ijk}a_i a_j a_k$ & (4.18) \\
\emph{the creation operators for neutrinos} & $c^3 \text{ by } \varepsilon^{ijk}c_i c_j c_k$ & (4.19) \\
\end{tabular}
\end{center}
\emph{where $a_i$ and $c_i$ now provide a basis for the fundamental representation of SU(3). Then the charged leptons and neutrinos are color singlets. }

\section*{The Knotted Electroweak Vectors} 

To achieve the required $U_a(1) \times U_b(1)$ invariance of the knotted Lagrangian (and the associated conservation of $t_3$ and $t_0$, or equivalently of the writhe and rotation charge), it is necessary to impose topological and empirical restrictions on the knotted vector bosons by which the knotted fermions interact as well as on the knotted fermions. \emph{For these electroweak vector fields we assume the $t=1$ of the standard model and therefore $j=3$ and $N=6$, in accord with \eqref{empirical} and \eqref{jN}} it follows \vspace{-0.2em} 
\begin{align*}
(j,m,m') = 3(t, -t_3, -t_0) \tag{4.10} \\
(j,m,m') = \frac{1}{2}(N,w,r+o) \tag{4.8}
\end{align*}
 that hold for the elementary fermion fields and \emph{that we now assume for the knotted vector fields as shown in Table 3.}
\begin{center} \vspace{-1.1em}
\begin{tabular}{r | r r r r l}
\multicolumn{6}{c}{\textbf{Table 3:} Electroweak Vectors $(j=3)$} \\
\hline \hline
 & $Q$ & $t$ & $t_3$ & $t_0$ & $D^{3t}_{-3t_3 - 3t_0}$ \\
 \hline
 $W^+$ & $e$ & 1 & $1$ & 0 & $D^3_{-3,0} \sim c^3d^3$ \\
 $W^-$ & $-e$ & 1 & $-1$ & 0 & $D^3_{3,0} \sim a^3b^3$ \\
 $W^3$ & 0 & 1 & 0 & 0 & $D^3_{0,0} \sim f_3(bc)$ \\
 \hline
\end{tabular}
\end{center}
The charged $W^+_{\mu}$ and $W^-_{\mu}$ are described by six preon monomials. The neutral vector $W^3_{\mu}$ is the superposition of four states of six preons given by \vspace{-0.2em}
\begin{align}
D^3_{00} = A(0,3)b^3c^3 + A(1,2)ab^2c^2d + A(2,1)a^2bcd^2 + A(3,0)a^3d^3 \tag{4.20} \vspace{-0.6em}
\end{align}
according to \eqref{long} which is reducible by the algebra (A) to a function of the neutral operator $bc$. Table 3 again illustrates the fact that the number of crossings equals the number of preons. To pass to a modified standard model in this table as well as to the dyon model, one must again replace $a$ and $c$ by $a_i$ and $c_i$ and antisymmetrize.

\break
\section{Graphical Representation of Corresponding Classical Structures}
The representation \eqref{long} of the four classical trefoils as composed of three overlapping preon loops is shown in Figure 1. In interpreting Figure 1, note that the two lobes of all the preon loops make opposite contributions to the rotation, $r$, so that the total rotation of each preon loop vanishes. When the three $a$-preons and $c$-preons are combined to form charged leptons and neutrinos, respectively, each of the three labelled circuits is counterclockwise and contributes $+1$ to the rotation while the single unlabeled and shared (overlapping) circuit is clockwise and contributes $-1$ to the rotation so that the total $r$ for both charged leptons and neutrinos is $+2$. \vspace{-0.3em}

For quarks the three labelled loops contribute $-1$ and the shared loop $+1$ so that $r=-2$.

In each case the three preons that form a lepton trefoil contribute their three negative rotation charges. The geometric and charge profile of the lepton trefoil is thus similar to the geometric and charge profile of a triatomic molecule composed of neutral atoms since the valence electronic charges of the atoms, which cancel the nuclear electronic charges of the atoms, are shared among the atoms to create the chemical binding of the molecule just as the negative rotation charges which cancel the positive rotation charges of the preons are shared among the preons to create the preon binding of the trefoils. There is a similar correspondence between quarks and antimolecules.

\newgeometry{bottom=2.5cm}
\begin{center}
\normalsize
\begin{tabular}{c | c} 
         \multicolumn{2}{c}{\textbf{Graphical Representation of Corresponding Classical Structures}} \\ [-0.54cm]
	\multicolumn{2}{c}{\textbf{Figure 1:} Preonic Structure of Elementary Fermions} \\ [-0.49cm]
	\multicolumn{2}{c}{$Q = -\frac{e}{6}(w+r+o)$, and $(j, m, m') = \frac{1}{2} (N, w, r+o)$} \\ [-0.25cm]
	\multicolumn{2}{c}{$D^j_{mm'} = D^{\frac{N}{2}}_{\frac{w}{2} \frac{r+o}{2}}$}\\
	\begin{tabular}{c c c}
		\multicolumn{3}{r}{ \ul{$(w, r, o)$}} \\ [-0.3cm]
		\multicolumn{3}{l}{Charged Leptons, $D^{3/2}_{\frac{3}{2} \frac{3}{2}} \sim a^3$} \\ [0.2cm]
		 \Large{$\varepsilon^{ijk}$} & $\hspace{17pt} a_j$ & \\ [0.4cm]
		 $\hspace{28pt} a_i$ & & $\hspace{-58pt} a_k$ \\ [-2.8cm]
		 \multicolumn{3}{l}{\xygraph{
!{0;/r1.3pc/:}
!{ \xoverh }
[u(1)] [l(0.5)]
!{\color{black} \xbendu }
[l(2)]
!{\vcap[2] =>}
!{\xbendd- \color{black}}
[d(1)] [r(0.5)]
!{\xunderh }
[d(0.5)]
!{ \xbendr}
[u(2)]
!{\hcap[2]=>}
[l(1)]
!{\xbendl- }
[l(1.5)] [d(0.25)]
!{\xbendl[0.5]}
[u(1)] [l(0.5)]
!{\xbendr[0.5] =<}
[u(1.75)] [l(2)]
!{ \xoverv}
[u(1.5)] [l(1)]
!{\color{black} \xbendr-}
[u(1)] [l(1)]
!{ \hcap[-2]=>}
[d(1)]
!{ \xbendl \color{black}}}} \\ [-1.6cm]
\multicolumn{3}{l}{ \hspace{33pt} \textcolor{red}{
\xygraph{
!{0;/r1.3pc/:}
[u(2.25)] [r(3.25)]
!{\xcapv[1] =>}
[l(2.75)] [u(.75)]
!{\xbendr[-1] =>}
[d(1)] [r(1.25)]
!{\xcaph[-1] =>}
}}}
 \\ [-0.7cm]
		 \multicolumn{3}{r}{\hspace{150pt}$(3,2,1)$}
		 
	\end{tabular}
	&
	\begin{tabular}{c c c}
		\multicolumn{3}{r}{\ul{$(w, r, o)$}} \\ [-0.3cm]
		\multicolumn{3}{l}{$a$-preons, $D^{1/2}_{\frac{1}{2} \frac{1}{2}}$} \\ [1.58cm]
		  & &\hspace{-63pt} $a_i$ \\ [-1.3cm]
		 \multicolumn{3}{l}{\xygraph{
!{0;/r1.3pc/:}
!{\xoverv=>}
[u(0.5)] [l(1)]
!{\xbendl}
[u(2)]
!{\hcap[-2]}
!{\xbendr-}
[r(1)]
!{\xbendr}
[u(2)]
!{\hcap[2]}
[l(1)]
!{\xbendl-}}} \\ [-1.2cm]
\multicolumn{3}{l}{\hspace{39pt} \textcolor{red}{\xygraph{
!{0;/r1.3pc/:}
[d(0.75)] [l(0.75)]
!{\xcaph[-1]=>}
}
}} \\ [-0.7cm]
		 \multicolumn{3}{r}{\hspace{150pt}$(1,0,1)$}
		 
	\end{tabular} \\ [-0.1cm] \hline
\begin{tabular}{c c c}
		\multicolumn{3}{l}{Neutrinos, $D^{3/2}_{-\frac{3}{2} \frac{3}{2}} \sim c^3$} \\ [0.2cm]
		 \Large{$\varepsilon^{ijk}$}& $\hspace{17pt} c_j$ & \\ [0.4cm]
		 $\hspace{28pt} c_i$ & & $\hspace{-58pt} c_k$ \\ [-2.8cm]
		 \multicolumn{3}{l}{\xygraph{
!{0;/r1.3pc/:}
!{\xunderh }
[u(1)] [l(0.5)]
!{\color{black} \xbendu}
[l(2)]
!{ \vcap[2]=>}
!{\xbendd- \color{black}}
[d(1)] [r(0.5)]
!{\xoverh}
[d(0.5)]
!{\xbendr}
[u(2)]
!{\hcap[2]=>}
[l(1)]
!{\xbendl-}
[l(1.5)] [d(0.25)]
!{ \xbendl[0.5]}
[u(1)] [l(0.5)]
!{\xbendr[0.5]=<}
[u(1.75)] [l(2)]
!{\xunderv}
[u(1.5)] [l(1)]
!{\color{black} \xbendr-}
[u(1)] [l(1)]
!{\hcap[-2]=>}
[d(1)]
!{\xbendl \color{black}}}} \\ [-1.5cm]
\multicolumn{3}{l}{\textcolor{red} {\hspace{32pt} \xygraph{
!{0;/r1.3pc/:}
[u(2.25)] [r(3.25)]
!{\xcapv[1] =<}
[l(2.75)] [u(.75)]
!{\xbendr[-1] =<}
[d(1)] [r(1.25)]
!{\xcaph[-1] =<}
}}} \\ [-0.7cm]
		 \multicolumn{3}{r}{\hspace{145pt}$(-3,2,1)$}
		 
	\end{tabular}
	&
	\begin{tabular}{c c c}
		\multicolumn{3}{l}{$c$-preons, $D^{1/2}_{-\frac{1}{2} \frac{1}{2}}$} \\ [1.58cm]
		  & &\hspace{-63pt} $c_i$ \\ [-1.3cm]
		 \multicolumn{3}{l}{\xygraph{
!{0;/r1.3pc/:}
!{\xunderv=>}
[u(0.5)] [l(1)]
!{\xbendl}
[u(2)]
!{\hcap[-2]}
!{\xbendr-}
[r(1)]
!{\xbendr}
[u(2)]
!{\hcap[2]}
[l(1)]
!{\xbendl-}}} \\ [-1.2cm]
\multicolumn{3}{l}{\textcolor{red}{\hspace{39pt} \xygraph{
!{0;/r1.3pc/:}
[d(0.75)] [l(0.75)]
!{\xcaph[-1]=<}
}}} \\ [-0.7cm]
		 \multicolumn{3}{r}{\hspace{145pt}$(-1,0,1)$}
		 
	\end{tabular} \\ [-0.1cm] \hline
\begin{tabular}{c c c}
		\multicolumn{3}{l}{$d$-quarks, $D^{3/2}_{\frac{3}{2} -\frac{1}{2}} \sim ab^2$} \\ [0.2cm]
		 & $\hspace{19pt} b$ & \\ [0.4cm]
		 $\hspace{28pt} a_i$ & & $\hspace{-56pt} b$ \\ [-2.8cm]
		 \multicolumn{3}{l}{\xygraph{
!{0;/r1.3pc/:}
!{\xoverh }
[u(1)] [l(0.5)]
!{\color{black} \xbendu}
[l(2)]
!{\vcap[2]=<}
!{\xbendd- \color{black}}
[d(1)] [r(0.5)]
!{ \xoverv}
[u(0.5)] [r(1)]
!{ \xbendr }
[u(2)]
!{\hcap[2]=<}
[l(1)]
!{\xbendl- }
[u(0.5)] [l(3)]
!{\xoverv}
[u(1.5)] [l(1)]
!{\color{black} \xbendr- }
[l(1)] [u(1)]
!{\hcap[-2]=<}
[d(1)]
!{\xbendl  \color{black}}
[r(2)] [u(0.5)]
!{\xcaph- =>}}} \\[-1.6cm]
\multicolumn{3}{l}{\hspace{42pt} \textcolor{red}{\xygraph{
!{0;/r1.3pc/:}
[u(1.75)] [l(1.5)]
!{\xbendd[-1]=<}
[u(0.75)] [r(1)]
!{\xbendl[-1]=<}
[d(1)] [l(2.25)]
!{\xcaph[-1]=>}
}}} \\ [-0.7cm]
		 \multicolumn{3}{r}{\hspace{140pt}$(3,-2,1)$}
		 
	\end{tabular}
	&
	\begin{tabular}{c c c}
		\multicolumn{3}{l}{$b$-preons, $D^{1/2}_{\frac{1}{2} -\frac{1}{2}}$} \\ [1.58cm]
		  & &\hspace{-63pt} $b$ \\ [-1.3cm]
		 \multicolumn{3}{l}{\xygraph{
!{0;/r1.3pc/:}
!{\xoverv=<}
[u(0.5)] [l(1)]
!{\xbendl}
[u(2)]
!{\hcap[-2]}
!{\xbendr-}
[r(1)]
!{\xbendr}
[u(2)]
!{\hcap[2]}
[l(1)]
!{\xbendl-}}} \\ [-1.5cm]
\multicolumn{3}{l}{\hspace{40pt} \textcolor{red}{\xygraph{
!{0;/r1.3pc/:}
[u(0.75)] [l(0.75)]
!{\xcaph[1]=>}
}}} \\ [-0.1cm]
		 \multicolumn{3}{r}{\hspace{145pt}$(1,0,-1)$}
		 
	\end{tabular} \\ [-0.1cm] \hline
\begin{tabular}{c c c}
		\multicolumn{3}{l}{$u$-quarks, $D^{3/2}_{-\frac{3}{2} -\frac{1}{2}} \sim cd^2$} \\ [0.2cm]
		 & $\hspace{19pt} d$ & \\ [0.4cm]
		 $\hspace{28pt} c_i$ & & $\hspace{-56pt} d$ \\ [-2.8cm]
		 \multicolumn{3}{l}{\xygraph{
!{0;/r1.3pc/:}
!{\xunderh }
[u(1)] [l(0.5)]
!{\color{black} \xbendu}
[l(2)]
!{\vcap[2] =<}
!{\xbendd- \color{black}}
[d(1)] [r(0.5)]
!{\xunderv}
[u(0.5)] [r(1)]
!{\xbendr}
[u(2)]
!{\hcap[2] =<}
[l(1)]
!{\xbendl-}
[u(0.5)] [l(3)]
!{\xunderv}
[u(1.5)] [l(1)]
!{\color{black} \xbendr-}
[l(1)] [u(1)]
!{\hcap[-2]=<}
[d(1)]
!{\xbendl \color{black}}
[r(2)] [u(0.5)]
!{\xcaph- =>}}} \\[-1.75cm]
\multicolumn{3}{l}{\hspace{44pt}\textcolor{red}{\xygraph{
!{0;/r1.3pc/:}
[u(1.75)] [l(1.5)]
!{\xbendd[-1]=>}
[u(0.75)] [r(1)]
!{\xbendl[-1]=>}
[d(1)] [l(2.25)]
!{\xcaph[-1]=<}
}}} \\ [-0.7cm]
		 \multicolumn{3}{r}{\hspace{140pt}$(-3,-2,1)$}
		 
	\end{tabular}
	&
	\begin{tabular}{c c c}
		\multicolumn{3}{l}{$d$-preons, $D^{1/2}_{-\frac{1}{2} -\frac{1}{2}}$} \\ [1.58cm]
		  & &\hspace{-63pt} $d$ \\ [-1.3cm]
		 \multicolumn{3}{l}{\xygraph{
!{0;/r1.3pc/:}
!{\xunderv=<}
[u(0.5)] [l(1)]
!{\xbendl}
[u(2)]
!{\hcap[-2]}
!{\xbendr-}
[r(1)]
!{\xbendr}
[u(2)]
!{\hcap[2]}
[l(1)]
!{\xbendl-}}} \\ [-1.5cm]
\multicolumn{3}{l}{\hspace{39pt} \textcolor{red}{\xygraph{
!{0;/r1.3pc/:}
[u(0.75)] [l(0.75)]
!{\xcaph[1]=<}
}}} \\ [-0.1cm]
		 \multicolumn{3}{r}{\hspace{140pt}$(-1,0,-1)$} \\		 
	\end{tabular} \\ [-0.43cm]
	\multicolumn{2}{c}{The clockwise and counterclockwise arrows are given opposite weights $(\mp 1)$ respectively.} \\ [-0.51cm]
	\multicolumn{2}{c}{The (rotation/writhe charge) is measured by the sum of the weighted (black/red) arrows.} \\
	[-0.51cm]
	\multicolumn{2}{c}{The central loops contribute oppositely to the rotation.}
\end{tabular}
\end{center}

\restoregeometry

\setlength{\baselineskip}{1.6\baselineskip}

\section{Presentation of the Model in the Preon Representation\cite{finkelstein15}}

The particles $(a,b,c,d)$ described in the following sections are either e or g preons and carry both e and g charges. The knot representation of $D^j_{mm'}$ by \eqref{long} as a function of $(a,b,c,d)$ and $(n_a,n_b,n_c,n_d)$ implies the constraints \eqref{n2j}, \eqref{n2m}, \eqref{n2m'} on the exponents in the following way:

\begin{IEEEeqnarray*}{+rCl+x*}
n_a+n_b+n_c+n_d & = & 2j & (3.11) \\
n_a+n_b - n_c - n_d & =  & 2m & (3.12) \\
n_a-n_b+n_c-n_d & = & 2m' . & (3.13)
\end{IEEEeqnarray*}
The two relations defining the quantum kinematics and giving physical meaning to $D^j_{mm'}$, namely the postulated \eqref{jN} and:
\begin{equation*}
(j,m,m')_q = \frac{1}{2}(N,w, r+o) \quad \text{field (flux loop) description} \tag{4.8}
\end{equation*}
and the semi-empirical \eqref{empirical}:
\begin{equation*}
(j,m,m')_q = 3(t, -t_3, -t_0)_L \quad \text{particle description} \tag{4.10}
\end{equation*}
imply two complementary interpretations of the relations \eqref{n2j}--\eqref{n2m'}. By \eqref{jN} one has a \textit{field} description $(N, w, \tilde{r})$ of the quantum state $(j, m, m')$ as follows
\begin{equation}
\left.
{\arraycolsep=1.2pt
\begin{array}{rl}
N &= n_a + n_b + n_c + n_d \\
w &= n_a + n_b - n_c - n_d  \\
\tilde{r} \equiv r+o &= n_a - n_b + n_c - n_d
\end{array}
}
\quad \right\} \text{field (flux loop) (N, w, $\tilde{r}$) description} \label{field (flux loop) description}
\end{equation}
In the last line of \eqref{field (flux loop) description}, where $\tilde{r} \equiv r+o$ and $o$ is a parity index, $\tilde{r}$ has been termed ``the quantum rotation," and $o$ the ``zero-point rotation." 

By \eqref{empirical} one has a "\textit{particle} description" $(t, t_3, t_0)$ of the same quantum state $(j, m, m')$.
\begin{equation}
\left.
{\arraycolsep=1.2pt
\begin{array}{rl}
t &= \frac{1}{6} (n_a + n_b +n_c +n_d)  \\
t_3 &= -\frac{1}{6}(n_a + n_b - n_c - n_d)  \\
t_0 &= -\frac{1}{6}(n_a - n_b + n_c - n_d) 
\end{array}
}
\quad \right\} \text{particle ($t$, $t_3$, $t_0$) description} \label{particle_description}
\end{equation}
\emph{In \eqref{particle_description}, $(t, t_3, t_0)$ are to be read as SLq(2) preon indices agreeing with standard SU(2) $\times$ U(1) notation} only \emph{at $j = \frac{3}{2}$. In general $t_3$ measures writhe charge, $t_0$ measures rotation hypercharge and $t$ measures the total preon population or the total number of crossings of the associated classical knot.}

\section{Particle--Field Complementarity in Preon Representation}
In the flux loop description equations \eqref{field (flux loop) description}, the numerical coefficients may be replaced by $(N_p, w_p, \tilde{r}_p)$ describing the preons as follows: \vspace{-0.5em} 
\begin{center}
\vspace{-0.5em}
\renewcommand{\tabcolsep}{3pt}
\begin{tabular}{l r r}
\begin{tabular}{l l r l}
 & $N = \sum_p n_p N_p$ & &\\ 
 & $ w = \sum_p n_p w_p $\hspace{-15pt} \hspace{-5pt} & & \hspace{10pt} where  \\
 & $\tilde{r} = \sum_p n_p \tilde{r}_p$ & & \hspace{50pt} \\
\end{tabular}
&
\begin{tabular}{r | r r r}
\hline \hline
$p$ & $N_p$ & $w_p$ & $\tilde{r}_p$ \\
\hline
$a$ & 1 & 1 & 1 \\
$b$ & 1 & 1 & $-1$ \\
$c$ & 1 & $-1$ & 1 \\
$d$ & 1 & $-1$ & $-1$ \\
\hline
\end{tabular}
& \hspace{20pt}
\begin{tabular}{c}
(7.1) \\
(7.2) \\
(7.3)
\end{tabular}
\end{tabular}
\vspace{0.5em}
\end{center} \vspace{-0.3em}

and in the particle description equations \eqref{particle_description}, the numerical coefficients may be replaced by $(t_p, t_{3_p}, t_{0_p})$ describing the preons as follows:
\begin{center}
\vspace{-0.5em}
\renewcommand{\tabcolsep}{3pt}
\begin{tabular}{l r r}
\begin{tabular}{l l r l}
$t = \sum_p n_p t_p$ & & & \hspace{50pt} \\
$ t_3 = \sum_p n_p t_{3_p} $  &\hspace{-15pt} \hspace{-5pt} & & \hspace{10pt} where \\
$ t_0 = \sum_p n_p t_{0_p} $  &&& \\
\end{tabular}
& \normalsize
\begin{tabular}{r | r r r }
\hline \hline
$p$ & $t_p$ & $t_{3_p}$ & $t_{0_p}$ \\
\hline
$a$ & $\frac{1}{6}$ & $-\frac{1}{6}$ & $-\frac{1}{6}$ \\
$b$ & $\frac{1}{6}$ & $-\frac{1}{6}$ & $\frac{1}{6}$ \\
$c$ & $\frac{1}{6}$ & $\frac{1}{6}$ & $-\frac{1}{6}$  \\
$d$ & $\frac{1}{6}$ & $\frac{1}{6}$ & $\frac{1}{6}$  \\
\hline
\end{tabular}
& \hspace{20pt}
\begin{tabular}{c}
(7.4) \\
(7.5) \\
(7.6)
\end{tabular}
\end{tabular}
\vspace{0.5em}
\end{center} \vspace{-0.3em}
\addtocounter{equation}{6}
In (7.1)-(7.6), $p$ is summed over $a, b, c, d$.

Since $r=0$ for preonic loops, $o$ plays the role of a quantum rotation for preons:
\begin{align}
\tilde{r}_p = r_p + o_p = o_p \qquad \qquad p = (a, b, c, d)
\end{align}
For the elementary fermions presently observed, we have set
\begin{align}
\tilde{r} = r + 1. 
\end{align}
\emph{The formal algebraic relations (7.1)--(7.6) express properties of the higher representations of the SLq(2) algebra as additive compositions of the fundamental representation.} \emph{The quantum state is now defined by $(n_a, n_b, n_c, n_d)$, the preon populations. It is as well still defined by $(j, m, m')$, the SLq(2) representation, and by the complementary knot $(N, w, \tilde{r})$ and particle descriptions, $(t, t_3, t_0)$}. All of these descriptions impose the same quantum kinematics.

\section{Interpretation of the Complementary Equations Continued}
There is also an alternative particle interpretation of the flux loop equations \eqref{field (flux loop) description}
\begin{align*}
N = n_a + n_b + n_c + n_d \tag{6.1$N$} \\
w = n_a + n_b - n_c - n_d \tag{6.1$w$} \\
\tilde{r} = n_a - n_b + n_c - n_d \tag{6.1$\tilde{r}$}
\end{align*}

Here the left-hand side with coordinates $(N,w,r)$ label a 2d-projected knot, and the right-hand side describes the preon population of the corresponding quantum state. 

\emph{Equation (6.1$N$) states that the number of crossings, $N$, equals the total number of preons, $N'$, as given by the right side of this equation. Since we assume that the preons are fermions, the knot describes a fermion or a boson depending on whether the number of crossings is odd or even.} Viewed as a knot, a fermion becomes a boson when the number of crossings is changed by attaching or removing a geometric curl
\begin{turn}{90}
\xygraph{
    !{0;/r0.75pc/:}
    !{\xunderh}
    [u l(0.75)]!{\xbendu}
    [l (1.5)]!{\vcap[1.5]}
    !{\xbendd-}}
\end{turn}
. This picture is consistent with the view of a curl as an opened preon loop, in turn viewed as a twisted loop
\begin{turn}{90}
\xygraph{
    !{0;/r0.75pc/:}
    !{\xunderh}
    [u l(0.75)]!{\xbendu}
    [l (1.5)]!{\vcap[1.5]}
    !{\xbendd-}
    [l] [d(0.5)]!{\xbendu-}
    [ld]!{\vcap[-1.5]}
    [u] [r(0.5)]!{\xbendd}
}
\end{turn}. Each counterclockwise or clockwise classical curl corresponds to a preon creation operator or antipreon creation operator respectively.
\newline

Since $a$ and $d$ are creation operators for antiparticles with opposite charge and hypercharge, while $b$ and $c$ are neutral antiparticles with opposite values of the hypercharge, we may introduce the preon numbers
\begin{align}
\nu_a &= n_a -n_d \\
\nu_b &= n_b - n_c
\end{align}
Then (6.1$w$) and (6.1$\tilde{r}$) may be rewritten in terms of preon numbers as
\begin{align}
&\nu_a + \nu_b = w \hspace{3pt} (= -6t_3) \\
&\nu_a - \nu_b = \tilde{r} \hspace{3pt} (=-6t_0)
\end{align}
By (8.3) and (8.4) the conservation of the preon numbers and of the charge and hypercharge is equivalent to the conservation of the writhe and rotation, which are topologically conserved at the 2d-classical level. In this respect, these quantum conservation laws for preon numbers correspond to the classical conservation laws for writhe and rotation.

Eqns. $(6.1N) - (6.1\tilde{r})$may also be interpreted directly in terms of Fig. 2 by describing the right-hand side of these equations as the possible populations of the conjectured preons at these crossings of Fig. 2 and interpreting the left-hand side as parameters of the binding field that links the 3 conjectured preons.
\newline

\section{Summary on the Measure of Charge by SU(2) $\times$ U(1) and by SLq(2)} 
The SU(2)$\times$U(1) measure of charge requires the assumption of fractional charges for the quarks. The SLq(2) measure requires the replacement of the fundamental charge $(e)$ for charged leptons by a new fundamental charge $(e/3)$ or $(g/3)$ for charged preons and then does not require fractional charges for quarks.\cite{shupe79}

The SLq(2), or $(j,m,m')$ measure, has a direct preon interpretation since $2j$ is the total number of preonic sources, while $2m$ and $2m'$ respectively measure the numbers of writhe and rotation sources of preonic charge.\cite{finkelstein15} Since $N$, $w$, and $r$ all measure the handedness of the source, charge is also measured by the chirality of the source.

If neutral unknotted flux tubes predated the particles, and the particles were initially formed by the knotting of unknotted flux tubes of energy-momentum, then the simplest fermions that could have formed with topological stability must have had three crossings and therefore three preons. \emph{The electric charge of the resultant trefoil or of any composite of preons would then be a measure of the chirality generated by the knotting of an original unknotted flux loop of energy-momentum.}

The total SLq(2) charge sums the signed two dimensional clockwise and counterclockwise turns that any energy-momentum current makes both at the crossings and in making a single circuit of the 2d-projected knot. This measure of charge, ``knot charge", which is suggested by the leptons and quarks, appears more fundamental than the electroweak isotopic measure that originated in the neutron-proton system, since it reduces the concept of charge to the chirality of the corresponding energy-momentum curve which may also be described as a SLq(2) chirality and in this way reduces charge to a topological concept similar to the way energy-momentum is geometrized by the curvature of spacetime.

\section{Possible Physical Interpretations of Corresponding Quantum States}

Since one may interpret the elements $(a,b,c,d)$ of the SLq(2) algebra as creation operators for either preonic particles or current loops, the $D^j_{mp}$ may be interpreted as a creation operator for a composite quantum particle composed of either preonic particles $(N', \nu_a, \nu_b)$ or current loops $(N, w, \tilde{r})$ . \emph{These two complementary views of the same particle may be reconciled as describing $N$-preon systems bound by a knotted field having $N$-crossings with the preons at the crossings as illustrated in Figure 2 for $N=3$.} In the limit where the three outside lobes become small or infinitesimal compared to the central circuit, the resultant structure will resemble a three particle system tied together by a string.
\begin{figure}[!tb]
\begin{center}
\normalsize
\begin{tabular}{c | c}
	\multicolumn{2}{c}{\textbf{Figure 2:} Leptons and Quarks Pictured as Three Preons Bound by a Trefoil Field} \\ [-0.0cm]
	\begin{tabular}{c c c}
		\multicolumn{3}{r}{ \ul{$(w, r, o)$}} \\ [-0.3cm]
		\multicolumn{3}{l}{Neutrinos, $D^{3/2}_{-\frac{3}{2} \frac{3}{2}} \sim c^3$} \\ [0.2cm]
		 & $\hspace{17pt} c_j$ & \\ [-.22 cm]
		 \multicolumn{3}{c}{\hspace{-1.655 cm}\vspace{0.22 cm}\textcolor{blue} {\scalebox{2}{\Huge .}}} \\ [-0.6cm]
		 $\hspace{28pt} c_i$ {\hspace{.33cm}\textcolor{blue} {\scalebox{2}{\Huge .}}} \hspace{-1cm} & & {\hspace{-2.65cm}\textcolor{blue} {\scalebox{2}{\Huge .}}} \hspace{2.15cm} $\hspace{-56pt} c_k$ \\ [-2.8cm]
		 \multicolumn{3}{l}{\xygraph{
!{0;/r1.3pc/:}
!{\xunderh }
[u(1)] [l(0.5)]
!{\xbendu}
[l(2)]
!{\vcap[2]=>}
!{\xbendd-}
[d(1)] [r(0.5)]
!{\xoverh}
[d(0.5)]
!{\xbendr}
[u(2)]
!{\hcap[2]=>}
[l(1)]
!{\xbendl-}
[l(1.5)] [d(0.25)]
!{\xbendl[0.5]}
[u(1)] [l(0.5)]
!{\xbendr[0.5]=<}
[u(1.75)] [l(2)]
!{\xunderv}
[u(1.5)] [l(1)]
!{\xbendr-}
[u(1)] [l(1)]
!{\hcap[-2]=>}
[d(1)]
!{\xbendl}}} \\[-1.5cm]
\multicolumn{3}{l}{\textcolor{red} {\hspace{32pt} \xygraph{
!{0;/r1.3pc/:}
[u(2.25)] [r(3.25)]
!{\xcapv[1] =<}
[l(2.75)] [u(.75)]
!{\xbendr[-1] =<}
[d(1)] [r(1.25)]
!{\xcaph[-1] =<}
}}} \\ [-0.7cm]
		 \multicolumn{3}{r}{\hspace{145pt}$(-3,2,1)$}
		 
	\end{tabular}
&
\begin{tabular}{c c c}
		\multicolumn{3}{r}{ \ul{$(w, r, o)$}} \\ [-0.3cm]
		\multicolumn{3}{l}{Charged Leptons, $D^{3/2}_{\frac{3}{2} \frac{3}{2}} \sim a^3$} \\ [0.2cm]
		 & $\hspace{17pt} a_j$ & \\ [-.22 cm]
		 \multicolumn{3}{c}{\hspace{-1.58 cm}\vspace{0.22 cm}\textcolor{blue} {\scalebox{2}{\Huge .}}} \\ [-0.6cm]
		 $\hspace{28pt} a_i$ {\hspace{.33cm}\textcolor{blue} {\scalebox{2}{\Huge .}}} \hspace{-1cm}  & & {\hspace{-2.65cm}\textcolor{blue} {\scalebox{2}{\Huge .}}} \hspace{2.15cm} $\hspace{-58pt} a_k$ \\ [-2.8cm]
		 \multicolumn{3}{l}{\xygraph{
!{0;/r1.3pc/:}
!{\xoverh }
[u(1)] [l(0.5)]
!{\xbendu}
[l(2)]
!{\vcap[2]=>}
!{\xbendd-}
[d(1)] [r(0.5)]
!{\xunderh}
[d(0.5)]
!{\xbendr }
[u(2)]
!{\hcap[2]=>}
[l(1)]
!{\xbendl-}
[l(1.5)] [d(0.25)]
!{\xbendl[0.5]}
[u(1)] [l(0.5)]
!{\xbendr[0.5]=<}
[u(1.75)] [l(2)]
!{\xoverv}
[u(1.5)] [l(1)]
!{\xbendr-}
[u(1)] [l(1)]
!{\hcap[-2]=>}
[d(1)]
!{\xbendl}}} \\ [-1.6cm]
\multicolumn{3}{l}{\hspace{33pt} \textcolor{red}{
\xygraph{
!{0;/r1.3pc/:}
[u(2.25)] [r(3.25)]
!{\xcapv[1] =>}
[l(2.75)] [u(.75)]
!{\xbendr[-1] =>}
[d(1)] [r(1.25)]
!{\xcaph[-1] =>}
}}} \\ [-0.7cm]
		 \multicolumn{3}{r}{\hspace{150pt}$(3,2,1)$}
		 
	\end{tabular} \\ [-0.1cm] \hline
	\begin{tabular}{c c c}
		\multicolumn{3}{l}{$d$-quarks, $D^{3/2}_{\frac{3}{2} -\frac{1}{2}} \sim ab^2$} \\ [0.2cm]
		 & $\hspace{19pt} b$ & \\ [-.22 cm]
		  \multicolumn{3}{c}{\hspace{-1.52 cm}\vspace{0.22 cm}\textcolor{blue} {\scalebox{2}{\Huge .}}} \\ [-0.6cm]
		 $\hspace{28pt} a_i$ {\hspace{.33cm}\textcolor{blue} {\scalebox{2}{\Huge .}}} \hspace{-1cm}  & & {\hspace{-2.65cm}\textcolor{blue} {\scalebox{2}{\Huge .}}} \hspace{2.15cm} $\hspace{-56pt} b$ \\ [-2.8cm]
		 \multicolumn{3}{l}{\xygraph{
!{0;/r1.3pc/:}
!{\xoverh }
[u(1)] [l(0.5)]
!{\xbendu}
[l(2)]
!{\vcap[2]=<}
!{\xbendd-}
[d(1)] [r(0.5)]
!{\xoverv}
[u(0.5)] [r(1)]
!{\xbendr}
[u(2)]
!{\hcap[2] =<}
[l(1)]
!{\xbendl-}
[u(0.5)] [l(3)]
!{\xoverv}
[u(1.5)] [l(1)]
!{\xbendr-}
[l(1)] [u(1)]
!{\hcap[-2]=<}
[d(1)]
!{\xbendl}
[r(2)] [u(0.5)]
!{\xcaph- =>}}} \\ [-1.6cm]
\multicolumn{3}{l}{\hspace{42pt} \textcolor{red}{\xygraph{
!{0;/r1.3pc/:}
[u(1.75)] [l(1.5)]
!{\xbendd[-1]=<}
[u(0.75)] [r(1)]
!{\xbendl[-1]=<}
[d(1)] [l(2.25)]
!{\xcaph[-1]=>}
}}} \\ [-0.7cm]
		 \multicolumn{3}{r}{\hspace{140pt}$(3,-2,1)$}
		 
	\end{tabular}
	&
\begin{tabular}{c c c}
		\multicolumn{3}{l}{$u$-quarks, $D^{3/2}_{-\frac{3}{2} -\frac{1}{2}} \sim cd^2$} \\ [0.2cm]
		 & $\hspace{19pt} d$ & \\ [-.22 cm]
		  \multicolumn{3}{c}{\hspace{-1.75 cm}\vspace{0.22 cm}\textcolor{blue} {\scalebox{2}{\Huge .}}} \\ [-0.6cm]
		 $\hspace{28pt} c_i$ {\hspace{.33cm}\textcolor{blue} {\scalebox{2}{\Huge .}}} \hspace{-1cm}  & & {\hspace{-2.9cm}\textcolor{blue} {\scalebox{2}{\Huge .}}} \hspace{2.15cm} $\hspace{-56pt} d$ \\ [-2.8cm]
		 \multicolumn{3}{l}{\xygraph{
!{0;/r1.3pc/:}
!{\xunderh }
[u(1)] [l(0.5)]
!{\xbendu}
[l(2)]
!{\vcap[2]=<}
!{\xbendd-}
[d(1)] [r(0.5)]
!{\xunderv}
[u(0.5)] [r(1)]
!{\xbendr}
[u(2)]
!{\hcap[2]=<}
[l(1)]
!{\xbendl-}
[u(0.5)] [l(3)]
!{\xunderv}
[u(1.5)] [l(1)]
!{\xbendr-}
[l(1)] [u(1)]
!{\hcap[-2]=<}
[d(1)]
!{\xbendl}
[r(2)] [u(0.5)]
!{\xcaph- =>}}} \\ [-1.75cm]
\multicolumn{3}{l}{\hspace{44pt}\textcolor{red}{\xygraph{
!{0;/r1.3pc/:}
[u(1.75)] [l(1.5)]
!{\xbendd[-1]=>}
[u(0.75)] [r(1)]
!{\xbendl[-1]=>}
[d(1)] [l(2.25)]
!{\xcaph[-1]=<}
}}} \\ [-0.7cm]
		 \multicolumn{3}{r}{\hspace{140pt}$(-3,-2,1)$}
		 
	\end{tabular} \\ [-0.2cm]
\multicolumn{2}{c}{The preons conjectured to be present at the crossings are suggested by the blue dots at the crossings} \\ [-0.4cm]
\multicolumn{2}{c}{ of the lepton-quark diagrams, or at the crossings of any diagram with more crossings.}
\end{tabular}
\end{center}
\end{figure}
\emph{The physical models suggested by Fig. 2 may be further studied with the aid of preon Lagrangians similar to that given in reference 3}. The Hamiltonians of these three body systems may be parametrized by degrees of freedom characterizing both the preons and the binding field that come from the \emph{form factors required by SLq(2) invariance}. The masses of the leptons, quarks, and binding quanta are determined by the eigenvalues of this Hamiltonian in terms of the parameters describing the constituent preons and energy-momentum flux loops. There is currently no experimental guidance at these conjectured energies. These three body systems are, however, familiar in different contexts, namely
\begin{center}
H$^3$ composed of one proton and two neutrons: $PN^2$ \\
$P$ composed of one down and two up quarks: $DU^2$ \\
$N$ composed of one up and two down quarks: $UD^2$ \\
\end{center}
which are similar to
$U$ as $cd^2$ and $D$ as $ab^2$,
where $U$ and $D$ are up and down quarks, presented as three binding preon states. These different realizations of energy-momentum and charge represent different expressions of curvature and chirality, and in particular as displayed by a closed loop of energy-momentum.

\section{Alternate Interpretation}
In the model suggested by Fig. 2 the parameters of the preons and the parameters of the current loops are to be understood as codetermined. On the other hand, in an alternative interpretation of complementarity, the hypothetical preons conjectured to be present in Figure 2 carry no independent degrees of freedom and may simply describe \emph{concentrations of energy and momentum at the crossings of the energy-momentum tube}. In this interpretation of complementarity, $(t, t_3, t_0)$ and $(N,w, \tilde{r})$ are just two ways of describing the same quantum trefoil of field. \emph{In this picture the preons are bound}, i.e. they do not appear as free particles. This view of the elementary particles as either non-singular lumps of field or as solitons has also been described as a unitary field theory\textsuperscript{10}.

\section{Lower Representations}
We have so far considered the states $j = 3, \frac{3}{2}, \frac{1}{2}$ representing electroweak vectors, leptons and quarks, and preons, respectively. We finally consider the states $j=1$ and $j=0$. Here we shall not examine the higher $j$ states.

In the adjoint representation $j=1$, the particles are the vector bosons by which the $j=\frac{1}{2}$ preons interact and there are two crossings. These vectors are different from the $j=3$ vectors by which the $j=\frac{3}{2}$ leptons and the $j=\frac{3}{2}$ quarks interact.
\bigskip

If $j=0$, the indices of the quantum knot are
\begin{align}
(j,m,m')_q = (0,0,0)
\end{align}
and by the rule \eqref{jN} for interpreting the knot indices on the left chiral fields
\begin{align}
\frac{1}{2}(N,w,\tilde{r}) = (j, m, m')_q &= (0, 0, 0) 
\end{align}
Then the $j=0$ quantum states correspond to classical loops with no crossings $(N=0)$ just as preon states correspond to classical twisted loops with one crossing. Since $N=0$, the $j=0$ states also have no preonic sources of charge and therefore no electroweak interaction.  \emph{It is possible that these }$j=0$ \emph{hypothetical quantum states are realized as (electroweak non-interacting) loops of field flux with} $w=0$, $\tilde{r}= r+o = 0$\emph{, and }$r = \pm1$, $o =\mp1$ \emph{ i.e. with the topological rotation }$r=\pm1$. The two states $(r, o) = (+1, -1)$ and $(-1, +1)$ are to be understood as quantum mechanically coupled.

If, as we are assuming, the leptons and quarks with $j= \frac{3}{2}$ correspond to 2d projections of knots with three crossings, and if the heavier preons with $j= \frac{1}{2}$ correspond to 2d projections of twisted loops with one crossing, then if the $j=0$ states correspond to 2d projections of simple loops with no crossings, one might ask if these particles with no electroweak interactions and which are smaller and heavier than the preons, are among the candidates for ``dark matter." If these $j=0$ particles predate the $j=\frac{1}{2}$ preons, one may refer to them as ``yons" as suggested by the term ``ylem" for primordial matter.

\section{Speculations about an earlier universe and dark matter} \vspace{-0.8em}
One may speculate about an earlier universe before leptons and quarks had appeared, when there was no charge, and when energy and momentum existed only in the SLq(2) $j=0$ neutral state as simple loop currents of gravitational energy-momentum. Then the gravitational attraction would bring some pairs of opposing loops close enough to permit the transition from two $j=0$ loops into two opposing $j=\frac{1}{2}$ twisted loops. A possible geometric scenario for the transformation of two simple loops of current (yons) with opposite rotations into two $j=\frac{1}{2}$ twisted loops of current (preons) is suggested in Fig. 3. Without attempting to formally implement this scenario, one notes according to Fig. 3 that the fusion of two yons may result in a doublet of preons as twisted loops, which might also qualify as Higgs particles.

\begin{center}
\normalsize
\begin{tabular}{c  c  c}
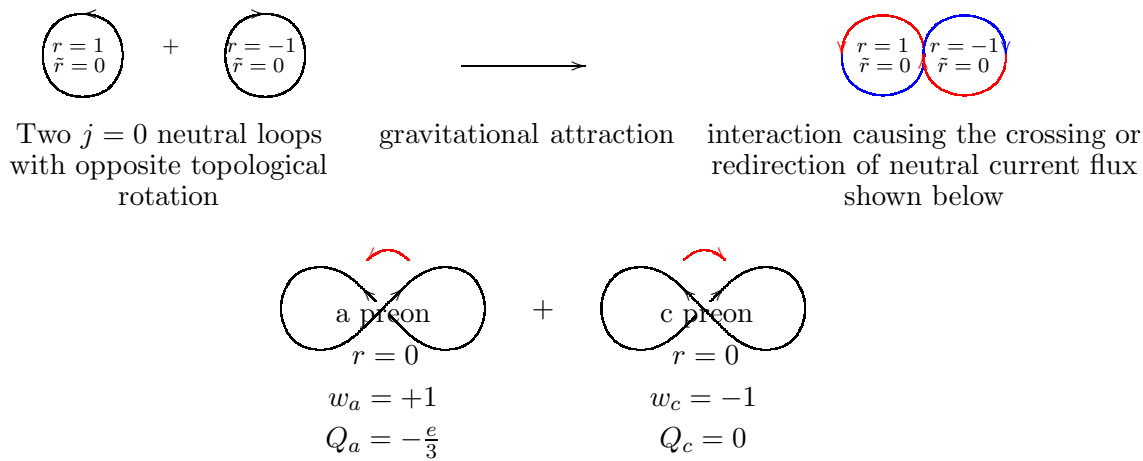
\multicolumn{3}{c}{ \textbf{Figure 3:} Creation of Preons as Twisted Loops} \\
\vspace{-3pt}
\xygraph{
!{0;/r1.3pc/:}
!{\hcap[2]=>}
!{\hcap[-2]}} \hspace{3pt} 
\xygraph{
!{0;/r1.3pc/:}
!{\hcap[2]}
!{ \hcap[-2]=< }} &
\xygraph{
!{0;/r1.3pc/:}
[d(1.25)]
!{\xcaph[3]=<@(0)}} & \hspace{-4pt}
\xygraph{
!{0;/r1.3pc/:}
[d(1)]
!{\color{red} \vcap[2]=> \color{blue}}
!{\vcap[-2]}} \hspace{-7pt} 
\xygraph{
!{0;/r1.3pc/:}
[d(1)]
!{\color{blue} \vcap[2]=<}
!{\color{red} \vcap[-2]=< \color{black}}}\\ [-1.7cm]
\hspace{4pt}\scriptsize{$r=1 \hspace{20pt}+ \hspace{16pt}r=-1$} & & \scriptsize{$\hspace{3pt}r=1 \hspace{8pt} r=-1$} \\ [-0.7cm]
\hspace{0pt}\scriptsize{$\tilde{r}=0 \hspace{32pt} \hspace{16pt}\tilde{r}=0$} & & \scriptsize{$\hspace{0pt}\tilde{r}=0 \hspace{10pt} \tilde{r}=0$} \\
Two $j=0$ neutral loops & gravitational attraction & interaction causing the crossing or  \\ [-0.55cm]
 with opposite topological & & redirection of neutral current flux \\ [-0.55cm]
  rotation & & shown below \\ [-0.55cm]
\end{tabular}
\end{center}
\begin{center}
\begin{tabular}{c c}
\textcolor{red}{\xygraph{
!{0;/r1.3pc/:}
[u(0.75)] [l(0.75)]
!{\xcaph[1]=>}
}} &
\textcolor{red}{\xygraph{
!{0;/r1.3pc/:}
[u(0.75)] [l(0.75)]
!{\xcaph[1]=<}
}} \\ [-1cm]
\xygraph{
!{0;/r1.3pc/:}
!{\xoverv=<}
[u(0.5)] [l(1)]
!{\xbendl}
[u(2)]
!{\hcap[-2]}
!{\xbendr-}
[r(1)]
!{\xbendr}
[u(2)]
!{\hcap[2]}
[l(1)]
!{\xbendl-}} &
\xygraph{
!{0;/r1.3pc/:}
!{\xunderv=<}
[u(0.5)] [l(1)]
!{\xbendl}
[u(2)]
!{\hcap[-2]}
!{\xbendr-}
[r(1)]
!{\xbendr}
[u(2)]
!{\hcap[2]}
[l(1)]
!{\xbendl-}} \\ [-1.1cm]
\multicolumn{2}{c}{+} \\ [-0.95cm]
\hspace{-95pt} & \hspace{-95pt} \\ [-0.9cm]
a preon & \hspace{-1pt} c preon \\ [-0.4cm]
\hspace{2pt}$r=0$ & \hspace{2pt}$r=0$ \\ [-0.4cm]
$w_a = +1$ & $w_c=-1$ \\ [-0.4cm]
$ Q_a = -\frac{e}{3}$ & $ Q_c = 0$  \\
\end{tabular}
\end{center}

In the scenario suggested by Figure 3 the opposing states are quantum mechanically entangled and may undergo gravitational exchange scattering. 

The $\binom{c}{a}$ doublet of Fig. 3 is similar to the Higgs doublet which is independently required by the mass term of the Lagrangian described in reference 3 to be a SLq(2) singlet $(j=0)$ and a SU(2) charge doublet $(t=\frac{1}{2})$. Since the Higgs mass contributes to the inertial mass, one expects a fundamental connection with the gravitational field at this point. 

If the fusion of the two yons yields a doublet of preons labelled as twisted loops, it could start a building-up process of elementary particles. 

If at an early cosmological time, only a fraction of the initial gas of quantum loops was converted to preons and these in turn led to a still smaller number of leptons and quarks, then most of the mass and energy of the universe would at the present time still reside in the dark loops while charge and current and visible mass would be confined to structures composed of leptons and quarks. \emph{In making experimental tests for particles of dark matter one might expect the SLq(2) $j=0$ dark loops to be different in mass than the dark neutrino trefoils where $j= \frac{3}{2}$, although both $j=0$ and $j=\frac{3}{2}$ would contribute to the dark matter.}

\section{Summary Comments on the Magnetoweak Phase}

The two sources of charge, $(\alpha_1, \alpha_2)$, carried by a dyonic particle normalize the SLq(2) algebra (A) by fixing the parameter 
\begin{equation}
q = \frac{\alpha_1}{\alpha_2}.
\end{equation}
We assume that the electroweak and magnetoweak sources of the model are related by the transposition of the coupling matrix $\varepsilon_q$ which may be interpreted as an expression of the parity conjugation satisfied by the electric and magnetic fields, and we define the two sources by 
\begin{equation}
\alpha_1 = \frac{e}{\sqrt{\hbar c}} \text{ and } \alpha_2 = \frac{g}{\sqrt{\hbar c}} \tag{14.2e}
\end{equation}
in the electroweak phase and 
\begin{equation}
\alpha_1 = \frac{g}{\sqrt{\hbar c}} \text{ and } \alpha_2 = \frac{e}{\sqrt{\hbar c}} \tag{14.2g}
\end{equation}
in the magnetoweak phase where $e$ and $g$ are the electroweak and magnetoweak charges. Then in the electroweak phase
\begin{equation}
q_e = \frac{e}{g} \tag{14.3e}
\end{equation}
and in the magnetoweak phase
\begin{equation}
q_g = \frac{g}{e} \tag{14.3g}
\end{equation}
and we assume $g>>e$.

\setcounter{equation}{3}
When the fundamental coupling matrix is also normalized by 
\begin{equation}
\text{det} \hspace{2pt} \varepsilon_q = 1
\end{equation}
then
\begin{equation}
eg = \hbar c
\end{equation}
and
\begin{equation}
q_e = \frac{e^2}{\hbar c} \tag{14.6e}
\end{equation}
while in the magnetoweak phase
\begin{equation}
q_g = \frac{\hbar c}{e^2}. \tag{14.6g}
\end{equation}
\setcounter{equation}{6}

To express the correspondence between electroweak and magnetoweak phases that suggests magnetic monopoles, we assume that the magnetoweak charges have a similar topological origin as the electroweak charges so that their sources are also describable as quantum trefoils. There are then magnetoweak as well as electroweak charged leptons, quarks and preons. All masses, both electroweak and magnetoweak, are partially fixed as Higgs masses by the eigenvalues of $\bar{D}^j_q (m,m') D^j_q (m,m')$ as follows\cite{finkelstein14a}
\begin{equation}
\left \langle n| \bar{D}^j_q (m,m') D^j_q (m,m') |n \right \rangle = f(q, \beta, n) \label{eigenD}
\end{equation}
where $|n \rangle$ are the eigenstates and $n$ labels the three empirical states of excitation of the leptons and quarks. Here $\beta$ is the value of $b$ on the ground state $|0 \rangle$. Since \eqref{eigenD} is a polynomial in $q$ and $\beta$ and of degree determined by $n$, the three lepton and quark masses may be parametrized by $q$, $\beta$ and $n$. The masses of the electroweak and magnetoweak charged preons, leptons and quarks, could then be vastly different since
\begin{align}
\frac{q_e}{q_g} & = \left (\frac{e^2}{\hbar c} \right )^2 \nonumber \\ 
& \approx \left ( \frac{1}{137} \right )^2 .
\end{align}

In the dyon interpretation of the SLq(2) algebra, one may speculate that the gluon binding is a form of magnetic binding and further that the dyon field passes from an early and high (cosmological) temperature g-phase to the current low temperature e-phase. 

One may speculate that the high temperature g-phase is composed of massive g-particles subject to correspondingly high attractive g-forces that can be opposed by a sufficiently high ambient temperature. In earlier historical periods, one supposes that the universe was much warmer and magnetic poles were more abundant and are possibly still observable in very deep probes of space. In this picture the current universe composed of leptons and quarks represents the e-phase of the dyon field.

\section*{ Acknowledgements}

I thank E. Abers, C. Cadavid, and J. Smit for comments on this paper, and S. Mackie for preparing this paper.

\end{document}